# Recent advances in wavefront shaping techniques for biomedical applications

**Hyeonseung Yu [a], Jongchan Park [a], KyoReh Lee [a], Jonghee Yoon [a], KyungDuk Kim [a], Shinwha Lee [b], YongKeun Park [a,\*]**

[a] *Department of Physics, Korea Advanced Institute of Science and Technology, Daejeon 305-701, Republic of Korea*
[b] *Department of Biological Sciences, Korea Advanced Institute of Science and Technology, Daejeon 305-701, Republic of Korea*
[\*] *E-mail: yk.park@kaist.ac.kr*

## 1. Introduction

Light transport in complex disordered media is a fundamental physical phenomenon, and it is significant in numerous applications including imaging through turbid layers, encryption, and quantum information processes [1]. At the interface between two media with different refractive indices, light refraction and reflection occur; when the interfaces become numerous and complicated such as paint layers, clouds, dense fog, the fine foam at the top of a glass of beer, and biological tissue, multiple light scattering events occur and it scrambles light paths in extremely disordered ways, which cause the objects behind or inside the scattering media to become opaque primarily due to multiple light scattering not absorption. Even though the resultant speckle patterns appear to exhibit highly disordered stochastic intensity distributions, multiple light scattering events are a deterministic process that can be precisely described using Maxwell's equations.

Recently, multiple research groups have exploited the deterministic nature of elastic light scattering and demonstrated the coherent control of multiple scattering in complex media using several wavefront shaping methods [3]. Measuring and controlling the wavefronts of speckle light fields in turbid media have been realized and accelerated through the advances of digital holographic techniques and the utilizations of wavefront modulators including spatial light modulators (SLMs).

Among the numerous potential applications of wavefront shaping techniques, the field of biomedical applications has been actively investigated. Imaging through opaque biological tissues has been a goal for a long time, because it may enable non-invasive optical diagnosis of diseases such as cancer. Although individual biological cells, which predominantly consist of water and proteins, are transparent in visible wavelengths, biological tissues that are composed of multiple layers of transparent biological cells are highly opaque, which results from the multiple light scattering. Wavefront shaping techniques for biomedical applications enable the suppression or circumvention of multiple light scattering with unique advantages. For example, wavefront shaping approaches do not require invasive processes such as the hydrogel electrophoresis in CLARITY [4] or incubation with an aqueous reagent in the Scale technique [5]. In addition, wavefront shaping approaches is applicable to a wide range of optical frequencies including visible wavelengths.

Currently, wavefront shaping techniques have become widely investigated in two methods in the field of biomedical optics. The first method optimizes the wavefront of an incident beam into a turbid medium via an SLM in order to generate an optical focus behind or inside the turbid medium [7-13]. The second method delivers optical information through a highly scattered layer via characterizing and exploiting multiple light scattering [14-16]. In order to provide more insight into how multiple light scattering can be exploited for various purposes, the recent progress in and techniques for wavefront shaping are summarized with an emphasis on their biomedical applications.

## 2. Exploiting multiple light scattering

### *2.1. Light scattering in biological tissues*

For biological tissue, refractive indices of components exhibit highly heterogeneous distributions and their lengths range from the nanometer scale to the millimeter scale. Therefore, light experiences multiple scattering upon propagating only a few tens of micrometers. In general, the light scattering properties of complex media are characterized using several optical parameters. The scattering coefficient $\mu_s$ and absorption coefficient $\mu_a$ describe a fractional decrease in the light intensity per unit distance traveled due to scattering and absorption, respectively. The scattering phase function $p(\theta)$ describes the angular dependence of scattering events, where $\theta$ is the deflection of angle of the scattered light [18, 19]. Scattering media, which exhibit anisotropic angular distributions of light scattering, are often described in terms of the anisotropy coefficient $g$, which defines the degree of forward scattering as the expectation value for the cosine of the scattering angle (i.e. $g = \langle \cos \theta \rangle$), and the reduced scattering coefficient $\mu'_s = \mu_s (1 - g)$.

Table 1 summarizes the values of $\mu_a$, $\mu'_s$, and $g$ of representative biological tissues, which have been compiled from Refs. [2, 6, 7, 17]. As seen in the table, light scattering is dominant over absorption in most biological tissues; the (reduced) scattering coefficients of the biological tissues are 10–100 times higher than the absorption coefficients, which implies that *in vivo* optical focusing and imaging are predominantly limited by light scattering not by absorption. The values of the anisotropy coefficient ($g$) are close to 1 for biological tissues, which indicates that forward scattering is dominant. The mean free path (MFP) and transport mean free path (TMFP) provide an intuitive understanding of the multiple light scattering in turbid media. The MFP, which is simply calculated as $1/\mu_s$, describes the average distance between two consecutive light scattering events. For turbid media with a high $g$, the TMFP, which is defined as $1/\mu'_s$, corresponds to the effective MFP considering forward light scattering. For most biological tissues, the TMFP is typically 0.1–1 mm.

The optical properties of biological tissues are wavelength dependent. The $\mu'_s$ decreases as the wavelength increases in many biological tissues [20]. Ultrasounds can obtain deep tissue macroscopic images of up to 10 cm [21], and near infrared (NIR) light can access up to several mm [22]. Visible light, which has been widely used as the light source of many microscopic techniques, only penetrates up to several hundred μm into tissues.

**Table 1**
Light scattering properties of representative biological tissues and non-biological samples.

|  | Reduced scattering coefficient $\mu'_s$ (cm$^{-1}$) | Anisotropy $g$ | Absorption coefficient $\mu_a$ (cm$^{-1}$) | Wavelength $\lambda$ (nm) |
| --- | --- | --- | --- | --- |
| Breast tissue [2] | 9.189 | 0.957 | 0.9 | 1000 |
| Liver tissue [2] | 8.112 | 0.952 | 0.5 | 1064 |
| Aorta [2] | 23.9 | 0.90 | 0.5 | 1064 |
| Myocardium [2] | 6.408 | 0.964 | 0.3 | 1064 |
| Enamel [6] | 2.4 | 0.96 | <1 | 632 |
| Cranial bone [17] | 19.48 | - | 0.11 | 800 |
| ZnO layers [7] | ~2000 | - | ~0 | 633 |

*2.2. Wavefront shaping techniques*

Until recently, it had been understood that multiple light scattering is a random process with an extremely high degree of freedom; thus, light transport in tissues has been considered using stochastic ensemble approaches such as the photon diffusion equation [23] and Monte Carlo simulations [24]. Multiple light scattering events were regarded as insurmountable barriers to imaging and manipulating light fields behind highly scattered layers until the first demonstration of coherent control of multiple scattering by Vellekoop *et al*. [7]. In their seminal work, an SLM, which is an array of liquid crystal pixels, was utilized to shape the wavefront of the light field impinging on the turbid layer, which generates a focus behind the layer. The basic principle is that at through optimizing the impinging wavefront, the optical phases of multiple light paths at a specific position behind the scattering layer, which were initially randomly distributed, become matched so that constructive interference is formed as a result of the linear relationship between the impinging and transmitting optical fields.

The principle of coherent control of multiple scattering is illustrated in Fig. 1. Using an optical lens, a light focus can be easily generated through an optically transparent layer with homogeneous distributions of refractive indices because the optical path lengths for each ray or optical phase delays of each spatial frequency components are precisely matched where the optical focus is formed (Fig. 1A).

However, when a light beam transmits through an inhomogeneous turbid layer, speckle patterns are generated after the turbid layer due to the spatial distribution of weak constructive and destructive interference conditions that results from the scrambling of the optical paths (Fig. 1B). Through optimizing the wavefront of the impinging beam, the relative phases of multiple light paths at the target point can be matched and thus an optical focus, i.e. a constructive interference pattern, can be generated (Fig. 1C). The optimization process is performed in an iterative manner (summarized in Ref. [25]). Focusing using a highly scattered media via wavefront shaping can also be understood using the time reversal property of light waves. When a beam that started from an optical focus passes through a scattering layer, it will have a complex wavefront and this is the same wavefront that needs to be located and optimized in order to generate an optical focus after passing through the scattering layer in the opposite direction.

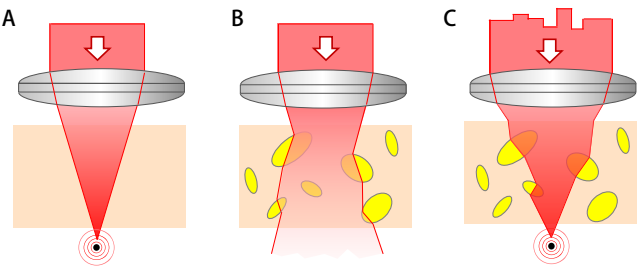

**Fig. 1.** (A) Focusing through an optically transparent layer using a lens; (B) the formation of speckle patterns in the presence of a turbid layer; and (C) through optimizing the wavefront of the impinging beam, the optical focus can be formed in the presence of a turbid layer.

This novel approach to generating an optical focus through a highly scattered layer has significant potential in biomedical optics, because the opaqueness of biological tissues primarily results from multiple light scattering with low absorptivity, which fundamentally limits optical imaging and treatment of diseases ranging from cancer diagnostics to photodynamic and photothermal therapy. This milestone experiment demonstrates that multiple light scattering can be systematically controlled through shaping the optical wavefronts. After the first demonstration of generating optical focuses through turbid media, this method has been combined with polarization [8, 26] and spectral selectivity [9, 27], as well as focusing broadband pulses [10, 28, 29].

*2.3. Scattering matrix approach*

The wavefront shaping technique has enabled the achievement of high controllability of light focusing over multiple scattering events; however, the control of light fields in turbid media is highly position dependent. Focusing on a newly targeted position requires an iterative optimization algorithm to be restarted from an unknown initial state. In contrast, in simple optical imaging systems using lenses, focusing on a newly targeted position is straightforward. In these linear spatially invariant systems, translating an optical output can be easily achieved through translating the input. However, light transport in complex media is a shift-variant system due to the extremely large degree of freedom with the characteristic length scale compatible with wavelengths of light.

Nonetheless, light transport in complex media is a linear process. Although the waves in complex media are scrambled upon propagation due to multiple scattering events and it appears unpredictable, it is a deterministic process: the relationship between the incident and outgoing waves can be determined precisely and it is unchanged upon the static state of the scattering sample. With a priori information about this relationship, optical information can be delivered through turbid layers and the desired wave field information can be calculated for incident beams in order to achieve the desired wave field after passing through the complex layer.

In general, the linear relationship between incident and scattered light fields is described using a scattering matrix [30]. A scattering matrix is composed of the reflection and transmission matrices, and each corresponds to the reflection and transmission of light through complex media, respectively. The measurements of a scattering matrix enable a complete description and manipulation of multiple scattering in complex media; however, its measurement is challenging for the following three reasons. First, due to the high degree of freedom, these matrices have vast numbers of elements [31]. Second, the sizes of the optical channels or bases of light waves become as small as the diffraction-limited spot size. Third, it requires the measurements of both the optical amplitude and phase information of speckle fields, and thus holographic or interferometric techniques are required

The first measurement of an optical transmission matrix (TM) was undertaken by Popoff *et al.* [32]. Using a common-path interferometry equipped with an SLM, the TM of an opaque layer constructed from ZnO nanoparticles was measured. Using the measured TM, the generation of optical foci through the opaque layer was demonstrated [32]. Furthermore, the image transmission through the opaque layer was demonstrated using the measured TMs [14].

The relationship between the TMs and wavefront shaping to focus through complex media is illustrated in Fig. 2. The measurement of the TM provides the complete information about the relationship between the input and output channels (Fig. 2A). In order to focus though a turbid layer, the wavefront shaping technique only utilizes a fraction of the TM, i.e. the information about the input channels that corresponds to a single output channel (Fig. 2B).

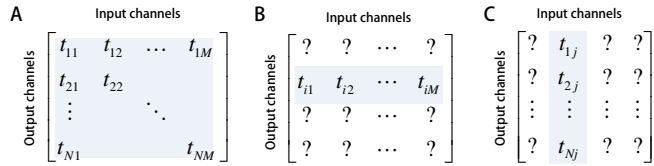

**Fig 2.** (A) The TM measurement acquires complete information about the relationship between the input and output channels. (B) The wavefront shaping technique for generating a focus accesses the information about the input channels corresponding to a single output channel (indicated by the shaded area). (C) The optical phase conjugation technique is achieved through addressing the information of all output channels corresponding to a single input channel.

Recently, a simple method to measure the TM using wavefront shaping was reported [33]. By modulating the wavefront of a beam impinging on complex medium and imaging the corresponding transmitted full-field speckle intensity patterns, a full TM of the turbid media can be precisely measured without using a complex interferometric optical system.

*2.4. Phase conjugation approach*

Optical phase conjugation, which is the transformation of a light field in a reversed propagation direction with an unchanged wavefront, has been used to suppress multiple scattering events and deliver imaging through turbid samples [34, 35]. After passing through a highly scattered layer, the incident optical field becomes a highly complex speckle field. When this speckle field is phase-conjugated and propagates in the reverse direction through the scattering layer, multiple scattering events are completely reversed in a time-reversal manner, which results in the formation of the original incident optical field after passing through the scattering layer, which can be achieved through addressing the information of all output channels corresponding to a single input channel (Fig. 2C).

Recently, digital optical phase conjugation (DOPC) has exhibited promising results in turbidity suppression in biological tissues [11-13, 36-40]. DOPC has two important steps: the measurement of a light field scattered from a scattering medium and the synthesis of the phase-conjugated scattered light field. The former can be achieved using digital holographic or quantitative phase imaging techniques [41-43], and the latter is performed using wavefront shaping techniques with SLMs.

Optical focusing through a turbid layer using DOPC was first demonstrate by Cui *et al*. [36]. In a subsequent study, DOPC was performed with live and excised rabbit tissues [37], in which the characteristic time scales for the decay of the DOPC signals were reported in a range of 2–30 sec due to the live motions in the tissues. The DOPC of fluorescence [38] and second-harmonic signals [39] have also been demonstrated. The single-shot DOPC of a full-field image requires extremely high precision in aligning optical paths, which was recently achieved through using an interferometer with a Sagnac-like ring design [40]. Recently, DOPC with ultrasound modulation has demonstrated potential for optical focusing inside turbid media [11] and high-resolution deep-tissue imaging [12, 13].

*2.5. Memory effect*

In many applications including optical imaging, it is often desirable to scan a focus in a two-dimensional or three-dimensional plane. However, generating multiple foci at different positions and scanning these reconstructed foci is time consuming and technically challenging. The optical memory effect enables the scanning of a constructed optical focus through a scattering layer.

When a coherent wave propagates through disordered media, the addition of a small phase gradient to an incident field (i.e. tilting of the incident field) results in the addition of the same phase gradient to a corresponding transmitted field [44]. Therefore, once an incident wavefront is optimized such that the optical focus to be generated at a target position behind a scattering layer, the generated focus can be translated through adding a small phase gradient to the optimized wavefront or through tilting the incident wavefront using a mirror or a SLM. Although the working angle range for the memory effect is narrow, memory effects have unique advantages in experiments. Within the working range of the memory effect, the wave transport in complex media can be regarded as a spatially invariant system.

Several interesting results have been reported using the memory effect [45-47]. Using wavefront shaping techniques and the memory effect, two dimensional images of fluorescent samples have been obtained through a scattering layer [45]. In this work, a spatial light modulator was used to optimize the impinging wavefront and a two-dimensional galvanometer was used to rotate the incident wavefront for the memory effect. The memory effect has also been used to scan a focus that was generated via optical phase conjugation [48] and to deliver optical images through scattering media via wavefront shaping and the memory effect [46].

*2.6. Overcoming the numerical aperture limit*

Utilizing these wavefront shaping techniques, scattering media can be used as optical lenses; generating and scanning the optical focus enables two-dimensional imaging of samples. In conventional imaging systems, the spatial resolving power is primarily determined by the numerical aperture (NA) of the optical system. Exploiting multiple light scattering via wavefront shaping techniques can enhance the effective NA of the optical imaging system because multiple light scattering events can produce high spatial frequency components. Vellekoop *et al*. first demonstrated the effective increase of NA in wavefront shaping and generated smaller optical foci than expected from the NA of the optical system [49]. This concept has been extended to two-dimensional imaging using the TM approach [50]. Later, the generation of optical foci smaller than 100 nm has been achieved through exploiting the effective increase of NA and using a high refractive index material [47].

Exploiting multiple light scattering in randomly distributed nanoparticles via wavefront shaping techniques enables the control and measurement of evanescent nearfields that are otherwise inaccessible with conventional optical lenses [51]. This concept was first demonstrated in microwave [52]. Recently, wavefront shaping techniques in multiple light scattering has been exploited in order to achieve subwavelength light focusing [53] and two-dimensional imaging [15].

# 3. Optical Implementations

*3.1. Wavefront modulator*

In order to control the wavefront of light fields, various types of wavefront modulators are commercially available and the choice of wavefront modulator depends on the aim of the experiments. Representative types of wavefront modulators [SLM, digital micro-mirror device (DMD), and deformable mirror (DM)] and their characteristics are summarized in Table 2.

SLMs using liquid crystals on silicon spatially modulate the wavefront of an optical field through controlling the alignment of the liquid crystal molecules [54]. In general, SLMs can control the phase or intensity of optical field depending on configurations. Using a SLM, both phase and amplitude can be independently controlled via phase-intensity cross modulation [55-57]. Phase-only SLM can precisely control the wavefront without changing the intensity or polarization of the light. The phases of optical fields from the surface of an SLM are controlled using the orientations of the liquid crystal molecules in the individual pixels, which are electrically reconfigurable. Commercial SLM devices are available in both transmission (liquid crystal display technology) and reflection types (liquid crystal on silicon technology). SLMs have high pixel resolutions, but their refresh rate is relatively slow in comparison with other types of wavefront modulators. Despite the capability of SLMs to control the phase, the high price of these SLMs prevents them from being widely applied in wavefront shaping techniques. Recently, a method to utilize commercial twisted nematic LCDs for full complex field modulation has been reported [55]. In comparison to SLMs, however, the efficiency of this method for wavefront shaping techniques is experimentally unfavorable due to its limited controllability of commercial twisted nematic LCDs [58]. More recently, light focusing through turbid media was demonstrated by controlling the polarization of incident light, in which a low-cost in-plane switching liquid crystal display (IPS-LCD) panel was utilized [59].

**Table 2**
Representative types of wavefront modulators and their characteristics.

|  | Spatial Light modulator (SLM) | Digital Micro-mirror Device (DMD) | Deformable Mirror (DM) |
| --- | --- | --- | --- |
| Working Principle | Electrically controlled liquid crystals arrays | Tilting of micro-mirror arrays | Piezoelectric arrays and flexible reflective surface |
| Pixel number | High (~ 800 × 600) | Ultra-high (~ 3,000 × 2,000) | Low (~ 200–5,000) |
| Control speed | Slow (10–100 Hz) | Ultra-fast (> kHz) | Ultra-fast (> kHz) |
| Cost (in USD) | ~30K | ~20K | ~ 50–100K |
| Diffraction efficiency | ~30–90% | < 50% | ~100% |
| Major manufacturers | Hamamatsu, Holoeye | Texas Instruments | Boston micromachines |

DMD contains arrays of micrometer-sized mirrors that can be tilted in on or off modes. In principle, DMDs are intensity modulators. Thus, the first demonstration of wavefront shaping using a DMD to generate a focus beyond the scattering layer was undertaken in a binary modulation mode [60]. However, DMDs can also be used as wavefront modulators; the optical phases can be precisely controlled using a DMD that employs an off-axis holography technique [61]. DMDs have a large number of pixels working at an extremely fast speed up to a few kHz. Fast wavefront shaping employing a DMD was demonstrated through diffusively moving scatters [61]. This high speed wavefront shaping is crucial for future *in vivo* biomedical applications. However, the diffraction efficiency of DMDs is low due to the grating nature of the subpixels.

DM utilizes the deformable reflective surface of which the shape can be controlled using arrayed piezoelectric actuators. The control speed of DMs is very high and the diffraction efficiency is almost 100%. However, the currently available DMs have a relatively small number of pixels and the device cost is high.

*3.2. Optimization algorithms for wavefront shaping*

For practical biomedical applications such as imaging and focusing through soft biological tissues, it is important to manipulate wavefronts with high speed and precision. Generating the focus behind complex media is a fundamental step in exploiting and controlling multiple light scattering. Several studies have investigated the analytically expression of the enhancement factor in order to generate a focus through turbid media using wavefront shaping techniques, which provide valuable information about the upper limit of the light scattering control [7, 60].

Based on the assumption that the TM entries, which represent the light transport through a scattering layer, are uncorrelated and follow the circular Gaussian distribution due to the central limit theorem, the intensity enhancement factors of the foci generated by wavefront shaping or binary amplitude modulation are analytically calculated as described below, respectively [7, 60]:

$$\eta_{phase} = \frac{\pi}{4}(N-1)+1,$$

$$\eta_{binary} = 1 + \frac{1}{\pi}\left(\frac{N}{2}-1\right),$$

where $\eta_{phase}$ and $\eta_{binary}$ denote the intensity enhancement factors when using a phase modulation method and a binary amplitude modulation method, respectively. $N$ is the number of modulating channels, which corresponds to the effective pixel numbers in the wavefront modulator. The reported experimental values for intensity enhancement factors were approximately 10–30% lower than the theoretical expectations, which might have resulted from the precision of SLMs and the decorrelation of speckle due to mechanical vibration, temperature variation, etc.

*3.3. Turbid samples*

Unprecedented controllability of light transport and focusing in turbid media is of particular interest to medical fields. For example, direct optical diagnostic and treatment of cancer without invasive procedures would be an important application field for wavefront shaping techniques in the future. However, the dynamic nature of biological tissues causes it to be extremely difficult to perform experiments for light control due to their short decorrelation time. In DOPC experiments with live tissues from a rabbit and a mouse, the focused signals exhibited a decay time ranging from 50 ms to 2.5 s depending on the degree of animal immobilization [37, 62]; however, *in vivo* studies of wavefront shaping techniques remain in their early stages. Therefore, most existing references have concentrated on understanding the potentials and limitations of wavefront shaping techniques in well controlled phantoms.

One of the most widely used scattering phantoms is a 2D layer composed of ZnO or $TiO_2$ nanoparticles [7]. Typically, these phantoms have MFPs that are comparable to the wavelengths of light. With a thickness of a few micrometers in the 2D layer, multiple light scattering events can occur. Commercial spray paints, consisting of $TiO_2$ nanoparticles with diameters smaller than wavelengths, have been used for creating scattering phantoms [15, 53]. Unlike a fragile 2D layer consisting of pure nanoparticles, commercial spray paints contain a fixative that makes the sample robust.

Scattering phantoms made from nanoparticles have several advantages: their optical parameters are well characterized and they are easy to fabricate. However, these scattering samples have MFPs (0.1–1 μm) that are three orders of magnitude shorter than those of biological tissues (0.1–1 mm). In order to create phantoms with controlled MFPs, the $TiO_2$ and PDMS composition [63] or the mixture of the intralipid and fibrin network [64] can be considered; these have been widely used in optical coherence tomography (OCT).

### 3.4. *The random matrix theory*

Multiple scattering of waves in disordered media has been investigated in electron transport theory over several decades. The same governing equations are applied to both electron transport and light transport, and the theory and experimental results in the electron transport have stimulated research in optical wavefront shaping techniques. One important branch of the theoretical studies is the random matrix theory (RMT), in which the correlations between the propagating modes inside the scattering media are examined based on the scattering matrix.

The interesting phenomenon of the "open channel" was predicted in the RMT; this phenomenon describes the perfect transmission through non-absorbing disordered media [30, 65]. A simple description of an open channel is that if turbid media only scatter light and not absorb it, light transport will occur in either forward scattering or backward scattering. In this condition, fully transmitting *open channels* or fully back scattering *closed channels* exist. The existence of open channels has been indirectly observed in light transport [66]. Recently, it was demonstrated that wavefront shaping can selectively excite open transport channels, resulting in an increase of up to 44% [67]. More recently, the observation of fully *open and closed channels* in a disordered waveguide has been reported by measuring the scattering matrix [68].

Optical open channels in turbid media are particularly interesting because in theory the full light transport inside biological tissues can be achieved via wavefront shaping regardless of the thickness of the tissues. However, attempts to address the open channels in light transport in turbid media have been stymied by fundamental limitations, particularly due to the inaccessibility of full optical modes. This inaccessibility is primarily inevitable for high spatial frequencies due to the NA of optics to illuminate and collect light to and from the scattering media in realistic optical experiments. Goetschy *et al*. theoretically investigated the realization of optical open channels and found that it was very difficult even with small losses of the optical information [69]. The inaccessibility to optical open channels has been experimentally demonstrated through monitoring the total transmission enhancement beyond the scattering media [70] and through measuring large optical TMs [71].

## 4. Biological applications

### 4.1. *Deep tissue imaging*

Over the past few decades, a significant breakthrough has been achieved in tissue imaging with the development of confocal microscopy [72], two-photon microscopy [73], photoacoustic microscopy [74], and optical coherence tomography [75]. However, the efforts to extend the penetration depth for deep tissue imaging, particularly in highly turbid tissues such as skin tissues, have been hindered by multiple light scattering. Recently, a significant breakthrough in deep tissue imaging has been achieved through using wavefront shaping techniques in the existing methods for imaging tissues.

Two-photon microscopy has been widely used for *in vivo* imaging of tissues and it has several advantages including high resolution three-dimensional imaging capability, the use of low energy photons, and more efficient photon detection in the scattering sample than that of confocal microscopy [73]. However, the aberrations caused by optical systems or samples significantly degrade both the optical resolution and penetration depth in two-photon microscopy. Recent research in two-photon microscopy has reported that adaptive optics can compensate the aberrations that are induced from weak scattering. Typically, the wavefront of reflected light from a sample surface is measured using a wavefront sensor, and then it is corrected using a wavefront modulator [76], [77]. Typically, a set of a Shack-Hartman sensors and a deformable mirror are used for the wavefront sensor and modulator, respectively. This adaptive optics approach corrects the aberrations that result from weakly scattered samples such as the retina. However, due to the limited degree of control, multiple scattering in turbid samples is not fully compensated.

Recently, wavefront shaping techniques have been used in deep tissue two-photon microscopy, and their potential for increasing the penetration depth in highly scattered samples has been demonstrated. Utilizing an SLM located at the input pupil of an objective lens, the wavefront of a beam impinging on the tissues was optimized in order to achieve diffraction-limited optical resolution [78] (Figs. 3A-C) and to increase the penetration depth in the two-photon microscopy [79]. Recent progress demonstrates that the aberration is effectively corrected through the interferometric focusing of light onto guide stars [80]; however, the requirement of the presence of a guide star inside the samples limits the range of applicable samples, particularly in *in vivo* focusing and imaging.

OCT is an established technique for non-invasive optical imaging of biological tissues. Based on the principle of low coherence interferometry, OCT provides the capability of optical sectioning [81]. OCT works in samples with more scattering than those for two-photon microscopy. However, the penetration depth of OCT is significantly reduced in highly scattered samples because OCT utilizes single back scattering signals as the image contrast; however, more than 99.9% of light reflected from biological tissues is repeatedly scattered [75].

In order to address the issue of multiple light scattering in OCT, several approaches have been reported recently. Fiolka *et al*. [82] demonstrated focusing through highly turbid biological tissues through shaping the impinging wavefront with backscattered light as a feedback signal. Our group has demonstrated that the use of a wavefront shaping technique can significantly enhance the penetration depth and signal-to-noise ratio in OCT signals (Figs. 3D and 3E) [83, 84]. Through optimizing the wavefront of an illumination beam using a DMD, the OCT signal at the target position, located using both the coherence and confocal gating, is increased, which results in the enhancement of the penetration depth in OCT up to 92% in highly scattered samples. Recently, Choi *et al*. presented the enhancement of light delivery into a scattering sample through measuring and exploiting its time-resolved reflection matrix [85]. More recently, optogenetic control through a highly scattering skull layer has been demonstrated *in vitro* using a wavefront shaping technique [86].

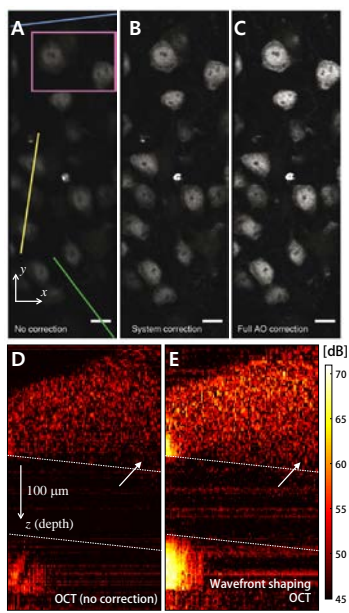

**Fig. 3. Wavefront shaping techniques for deep tissue imaging.** (A-C) The aberration-corrected two-photon microscopy: (A) image of a mouse brain slice without correction; (B) image with system aberration correction; and (C) the sample with induced aberration is further optimized resulting in an enhanced signal and resolution. (D-E) Wavefront-shaping OCT: B-line scan the OCT image of a highly scattered phantom (D) without the wavefront shaping technique and (E) with the wavefront shaping technique. The white arrows indicate the features seen in (E) that are not seen in (D). (A-C) and (D-E) are modified from Ref. [78] and Ref. [84] with permissions, respectively.

*4.2. Ultrasound-assisted light focusing inside turbid media*

Focusing and imaging inside biological tissues is one of the ultimate goals of wavefront shaping techniques due to their potential applications in non-invasive diagnosis and treatment of cancers. Although early works using wavefront shaping and TM approaches have demonstrated focusing and imaging *through* scattering samples, the field has further advanced to employ ultrasound modulation in order to focusing *inside* the scattering samples. Ultrasound focusing can function as a virtual guide star inside biological tissues because the ultrasound frequency is more immune to multiple scattering than the optical frequency, and light passing the ultrasound focus can be identified using the frequency shift.

Recently, the modulation of scattered light via ultrasound focusing has been exploited in order to filter out the scattered light that originated from the desired spot inside the medium, which is located via ultrasound focusing [11] (Figs. 4A and 4B). This technique exploits the facts that ultrasound wave can generate a focus deeper than optical wave inside complex media, and the scattered light that originated from an ultrasound focal volume can be frequency shifted due to acoustic modulation. Then, the acoustically modulated scattered light field can be selectively recorded via heterodyne interferometry, because scattered light that are only originated from the ultrasound focus have the shifted frequency, resulting into interference. This scattered light field originated from the ultrasound focus is then replayed to the original location of the ultrasound focus via the optical phase conjugation. In this work, a photorefractive crystal was used to record the transmitted optical field [11]. Later, the limitation of the low optical power of a reconstructing field in a photorefractive crystal has been considered using DOPC [12, 13]. In early works, the size of the ultrasound assisted light focusing inside turbid media was limited by the size of the ultrasound focus, which is significantly larger than that of the light focus, and this results in a significant restriction in the focusing and imaging resolutions. More recently, this spatial resolution barrier has been overcome through exploiting iterative sound-light interactions [87] and encoding individual spatial modes inside the scattering sample [88]. Recently, a method to measure a photoacoustic transmission matrix was reported, from which selective light focusing through diffusive samples has been demonstrated [89].

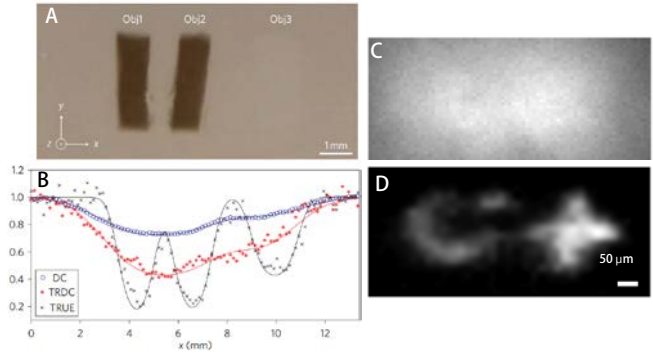

**Fig. 4. Ultrasound assisted light focusing.** (A-B) TRUE imaging with photorefractive crystal recording: (A) two absorbing objects (Obj1 and Obj2) and one scattering object (Obj3), and (B) reconstructed imaging using the TRUE method and comparisons with other modalities. (C-D) TRUE imaging with digital optical phase conjugation: (C) CIT feature patterned with quantum dots and (D) image acquired via TRUE imaging. (A-B) and (C-D) are modified from References [11] and [12], respectively, with permissions.

In addition, photoacoustic signal was utilized in a wavefront shaping technique in order to guide light into an absorbing sample embedded in optically diffusive media [90]. More recently, nonlinear photoacoustically guided wavefront shaping has been reported; dual-pulse excitation generates the Grueneisen relaxation effect, which was then used as feedback in iterative wavefront shaping in order to achieve optical diffraction-limited focusing in scattering media [91].

It could be envisioned that there are a number of interesting applications of light focusing inside turbid media, e.g. three-dimensional *in vivo* fluorescence imaging inside brain tissues, effective photodynamic therapy, and noninvasive photo-stimulation of cells and tissues. However, several technical innovations should be followed in order to achieve these goals. For example, the current optical implementation for DOPC requires extremely sensitive and complicated alignments. Furthermore, the intensity of ultrasound-modulated light is extremely low compared with unmodulated scattered light; thus, the long image acquisition time for existing setups should be improved for practical biological imaging.

### 4.3. Fiber-based Imaging

Endoscopic imaging, which is essential for modern gastrointestinal diagnosis and surgery, has evolved over the past few decades to enhance the image resolution and to employ multi-mode imaging capabilities [92]. However, the current endoscopic imaging techniques use a thick bundle of flexible glass fibers (typically 100,000 individual fibers packed in a 1.5 mm diameter bundle), and there have been efforts to build a thin optical image guiding fiber for endoscopic applications without reducing the resolution.

Recently, a significant breakthrough in endoscopic imaging has been achieved through exploiting wavefront shaping techniques in multi-mode optical fiber (MMF). A typical single MMF with a thickness of a few hundreds of micrometers can support tens of thousands of optical modes, and in principle the MMF can be utilized for ultra slim endoscopic imaging fibers. However, MMFs behave as turbid media, and the transmitted light does not form imaging directly, but rather it generates speckle patterns for coherent illumination. Several approaches have demonstrated that clever uses of wavefront shaping techniques can transform MMFs into endoscopic imaging fibers.

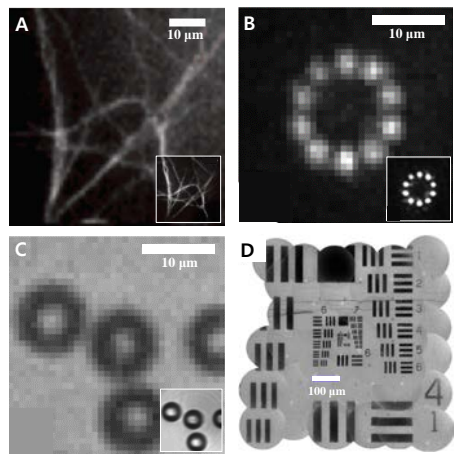

**Fig. 5. Image transmissions through single multimode fibers (MMF)** (A-C) The images through single MMFs using DOPC. (insets) original images. (D) The image of a USAF target obtained with a MMF employing the TM approach. Stitched images are obtained by moving the MMF. (A),(B-C),(D) are modified from Reference [93], [94] and [95] respectively, with permissions.

Holographic optical phase conjugation for image transmissions through single MMFs was proposed by Yariv in 1976 [96] and demonstrated by the Silberberg group in 1983 [97]. The first DOPC approach for imaging through MMFs was demonstrated by the Psaltis group in 2012 [98], in which one-to-one mapping of optical foci through an MMF was demonstrated using a SLM to compensate the modal dispersion via the phase conjugation method. Using this method, the subcellular images of fluorescently stained neuronal cells were obtained in a single MMF confocal microscopy [93] (Fig. 5A). The Dholakia group demonstrated the delivery of optical images through single MMFs, utilizing an SLM to decode the scrambled light transmitting an MMF and to recover the original image of a sample located in front of the MMF [94] (Figs. 5B and 5C). Using the TM approach, Choi *et al*. demonstrated wide-field endoscopic imaging through single MMFs without using a scanning device [95]. More recently, practical techniques towards endoscopic imaging using single MMFs have been reported including the enhancement of the image acquisition speed using DMD [99] and GPU [100], which is a compensation method for fiber bending modal dispersion [101].

*4.4. Wide-field imaging using memory effect*

After generating a single focus behind a scattering layer, it is possible to scan the generated focus through tilting the impinging wavefront toward the scattering layer, which exploits the optical memory effect [44]. Several applications of wide-field imaging behind scattering mediums have been demonstrated using the memory effect [45, 48, 102]. For example, using a raster scanning generated focus behind a scattering layer, fluorescent images can be measured (Fig. 6A) [45, 102]. This memory effect has also been employed to demonstrate imaging around corners with scattered incoherent light [46]. However, scanning a focus behind scattering media using the memory effect has a limited scanning area.

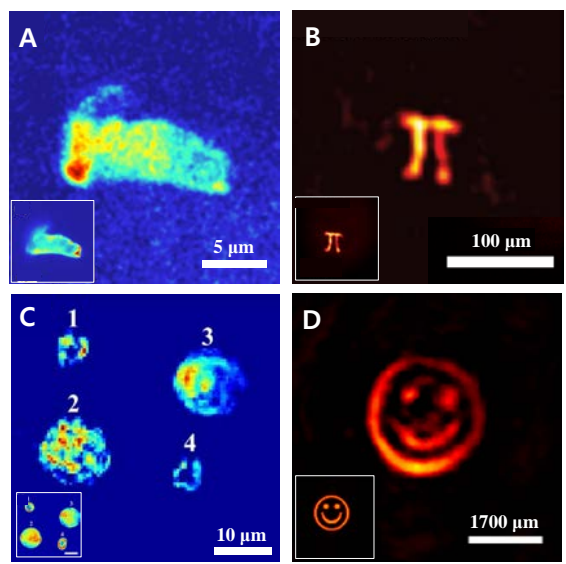

**Fig. 6. Wide field imaging using the memory effect.** (A) Scanning image of a dense cluster of 200 nm diameter fluorescent beads. (B) A 'π' shaped fluorescent object is retrieved using the autocorrelation method. (C) Eosin-Y stained blood cells are imaged using the autocorrelation method. (D) A single-shot fluorescence image through scattering media is obtained using a cell phone camera. (A-D) are modified from References [45], [16], [103], and [104], respectively, with permissions

Recently, the Mosk group demonstrated non-invasive imaging of a fluorescent object that was completely hidden behind a scattering layer [16]. Due to the memory effect, the angular autocorrelation of the measured diffusive fluorescence after the scattering layer was related to the autocorrelation of the object. From the angular autocorrelation of the fluorescence, which can be easily measured using a conventional optical imaging system, a high-resolution fluorescence image of the hidden object was retrieved via iterative Gerchberg-Saxton algorithm implementations (Fig. 6B). More recently, this method has been expanded to the imaging of biological cells (Eosin-Y stained blood cells in Fig. 6C) [103] and single-shot non-invasive imaging through turbid media using a smart-phone camera (Fig. 6D) [104]

## 5. Conclusion

The research work reviewed here indicates that various optical wavefront-shaping techniques may have an important function in overcoming multiple light scattering issues in complex biological tissues, which can have a critical impact on the future diagnosis and treatment of human diseases including cancer.

The uses of wavefront shaping techniques for biomedical applications have not yet been fully explored. This new emerging technology may grow quickly and confront important issues in imaging biological cells and tissues in opaque skin layers, and diagnosing and treating diverse pathological conditions. New techniques are particularly needed to enable a faster, and more compact and robust wavefront shaping for real-time *in vivo* applications. For example, the current DOPC technique requires extremely sensitive optical alignment in order to match the two planes of a holographic sensor and an actuator with sub-pixel accuracy [40]; ultrasound-assisted DOPC techniques require significant amounts of time for wide-field imaging. Furthermore, this emerging field of research can be integrated with optical trapping techniques in order to image and manipulate biological samples *in vivo*, as demonstrated in the proof-of-principle experiments *in vitro* [105].

Furthermore, the approaches in wavefront shaping technique can also benefit from combination with existing imaging modalities including multiphoton microscopy [106], optical coherence tomography [81], super resolution nanoscopy [107-110], and so on. Among these modalities, quantitative phase imaging can be readily integrated with wavefront shaping techniques because quantitative phase imaging directly measures the wavefront information of biological cells and tissues [42, 43, 111]. *In vivo* three-dimensional imaging of biological samples [112, 113], optical accessing of cellular drymass [114], and optical measurement of mechanical properties of biological cells [115-119] can be enabled *in vivo* through intact biological tissue layers by effectively suppressing multiple light scattering using wavefront shaping techniques.

In addition, wavefront shaping techniques can determine direct and potentially important applications in the diverse fields of biological and medical sciences. For example, *in vivo* imaging of blood cells flowing through skin tissue might have significant impacts in hematology [103] and direct counting of white blood cells or assessing hemoglobin content in red blood cells would be possible without requiring blood collection procedures [120-123]. Non-invasive optical detecting of rare circulating tumor cells or pathogenic bacteria could also be addressed in the near future [124, 125].

The recent developments in optical sciences have been effectively transferred from bench top to bedside as a result of close interdisciplinary collaboration; wide applications and utilizations of optical coherence tomography is a common example [81]. In order to successfully translate this exciting field of research to biological and medical applications, it is crucial to develop interdisciplinary collaboration among clinicians, biologists, engineers, and physicists, and then solve important issues in biomedical applications with simple, robust, and effective techniques. Considering the recent exponential growth of the field, we are optimistic that optical wavefront shaping techniques will have important functions in the study, diagnosis, and assessment of several diseases in the near future.

## Acknowledgements

This work was supported by KAIST-Khalifar University Project, APCTP, and National Research Foundation (NRF) of Korea, the Korean Ministry of Education, Science and Technology (MEST), and National Research Foundation of Korea (2012R1A1A1009082, 2012-M3C1A1-048860, 2013R1A1A3011886, 2013M3C1A3063046, 2013K1A3A1A09076135, 2014M3C1A3052537, 2014K1A3A1A09063027).